\documentclass[journal,11pt,onecolumn]{IEEEtran}
\usepackage{amsmath}
\usepackage{amsfonts}
\usepackage{amsthm}
\usepackage{dsfont}
\usepackage{epsfig}
\usepackage{upgreek}
\usepackage{graphicx}
\usepackage{caption}
\usepackage{setspace}

\theoremstyle{definition}

\newcommand{\Sm}{\mbox{${\bf S}$}}

\newcommand{\Mm}{\mbox{${\bf M}$}}
\newcommand{\Dm}{\mbox{${\bf D}$}}
\newcommand{\Dmi}{\mbox{${\bf D}^{\mbox{\tiny -1}}$}}

\newcommand{\Em}{\mbox{${\bf E}$}}
\newcommand{\Emi}{\mbox{${\bf E}^{\mbox{\tiny -1}}$}}

\begin{document}

\title{{A Scale-Consistent Approach for\\ Recommender Systems} }       
\author{{\Large Jeffrey Uhlmann}\\
Dept. of Electrical Engineering and Computer Science\\
University of Missouri-Columbia}

\date{}          
\maketitle
\thispagestyle{empty}


\begin{abstract}
In this paper we propose and develop a relatively simple and efficient 
matrix-completion approach 
for estimating unknown elements of a user-rating matrix in the 
context of a recommender system (RS). The critical theoretical property
of the method is its consistency with respect to arbitrary 
units implicitly adopted by different users to construct their
quantitative ratings of products. It is argued that this property
is needed for robust performance accuracy across a broad spectrum
of RS application domains.\\
{\tiny ~}
\end{abstract}

\section{Introduction}

Given a matrix (or higher tensor composition) in which the value associated with each element 
either represents a user's score/ranking of a particular item/product or is unfilled, a {\em recommender system} 
(RS) is intended to use information from existing values to estimate a value for each unfilled element, 
i.e., to estimate how a user is likely to score a particular product that has not yet been evaluated, in order 
to determine whether the item should be recommended to the user (\cite{ado,jannach,lakio,resnick}). 
Because this matrix-completion problem is
inherently ill-defined, certain assumptions must be made based on a premise that users who give 
similar scores to a common set of products are likely to give similar scores to other products. 
Unfortunately, this premise is far too simplistic to capture the complex multivariate structures
underpinning human interests and preferences.

To appreciate the RS challenge, consider a pair of users {\em Alice} and {\em Bob} of the same
age and with identical backgrounds (e.g., geographical, cultural, socio-economic, etc.), and then attempt
to explain the source of differences in their scores for a given set of products. A first potential source 
of difference arises from the fact that terms like {\em score} and {\em rating} are not rigorously 
defined, i.e., Alice and Bob are likely to adopt different ``units of quality'' when making their respective
assessments. These implicit units introduce scale factors that affect their distribution of scores even 
when they are forced to convert to a fixed scale, e.g., 0 to 10 or 1 to 100. Moreover, they are likely
to apply different units, or gradations of discrimination, to different product types, e.g., toothpaste
versus jackets. Their relative scores are also likely to reflect distinct personal preferences, e.g., horror
films versus romantic comedies. For example, Bob may enjoy horror films and apply a scale
in which his favorite horror films receive high scores while other horror films receive
low scores, whereas Alice may strongly dislike horror films and give the lowest possible score
to every film in that genre. In fact, a user may even consciously recognize that he is
applying different units of quality/preference when giving the same high score to, e.g., the Orson 
Welles film ``Citizen Kane'' and Bruce Lee's ``Enter the Dragon'' -- two films from completely
different genres with almost entirely disjoint attributes of appeal.

In this paper we propose use of a scale-consistent matrix transformation to permit unrated
elements of a matrix to be estimated in a manner that is robust to scale factors associated
with the scores of different users and with respect to scale factors associated with user ratings
of distinct product types or genres. For example, if Alice tends to give a 20\% higher score to
products in a particular category relative to Bob, then estimates/predictions of their respective 
scores for a new product in that category should reflect that 20\% factor.

\section{Diagonal Matrix Scalings}

To achieve unit-scale robustness in recommender system applications it is necessary to estimate
unrated values in a way that is as insensitive as possible to scale-factor differences in scores
given by different users across all products and with respect to scale-factor differences in scores 
for different product types/genres as received from across all users. To achieve this it is necessary to
reduce the given rating matrix to a unique scale-invariant form from which the estimates can then
be computed. More specifically, for a given $m\times n$ nonnegative matrix $\Mm$ a  
diagonal scaling is required that produces a {\em unique} matrix $\Mm'$ as
\begin{equation}
    \Mm' ~=~ \Dm\cdot\Mm\cdot\Em
\end{equation}
where $\Dm$ and $\Em$ are nonnegative diagonal matrices. A wide variety of scaling methods
have been investigated in the case of square $\Mm$ for the purpose of improving its condition
number as a precursor (preconditioning step) to performing a numerical linear algebra operation
that is highly sensitive to that condition number. Most methods scale only the rows or the columns,
and the ones that do scale both are not generally unique, i.e., very different scalings ($\Mm'$) may 
yield the same optimal condition number. 

An approach that does yield a unique result for nonnegative $\Mm$ is the 
{\em Sinkhorn scaling} (\cite{sink64,sink67}). 
The Sinkhorn scaling is achieved by iteratively scaling the rows to have unit sum, then the columns, etc., 
to convergence. Sinkhorn showed that this algorithm does in fact always converge to a unique result with
unit row and column sums.  However, that result is highly sensitive to the distribution of zeros in the
matrix, e.g., if applied to a triangular matrix the process will drive all nondiagonal values to zero with
no ability to define finite row and column ($\Dm$ and $\Em$) scalings. 

A less well known matrix scaling was defined by Rothblum and Zenios \cite{rz92} and can be used
to produce a scaled matrix with the property that the product of the nonzero elements in each row
and column is 1. As pointed out in \cite{simax}, the provable uniqueness of this scaling is 
analytically more important than the properties (e.g., condition number) of the result as it allows,
e.g., the derivation of a scale-consistent generalized matrix inverse that is required in many
practical engineering and robotics applications (\cite{simax,BoZ2,zhang}). 
As will be shown in the next section, it is also what 
is needed for RS robustness.

\section{The Algorithm}

Using the Rothblum \& Zenios (RZ) algorithm it is possible to obtain a
unique scaled matrix $\Sm$ from nonnegative matrix $\Mm$ as 
\begin{equation}
   \Sm ~=~ \Dm\cdot\Mm\cdot\Em
\end{equation}
where $\Dm$ and $\Em$ are nonnegative diagonals and the
product of the nonzero elements of each row and column of $\Sm$
is 1.

It should be noted that zero elements are invariant in the RZ decomposition, 
so if unfilled entries in $\Mm$ are taken to be zero then they will also be
zero in $\Sm$. Maintaining the distinction between zeros representing
scores versus zeros representing unfilled/unknown values, we replace the 
latter with $1$s in $\Sm$ based on the rationale that $1$ is the only nonzero
value that can be used that will preserve the unit-product property of $\Sm$.
Denoting the result as $\Sm'$, a filled matrix $\Mm'$ can be obtained by
inverting the original scaling as
\begin{equation}
   \Mm' ~=~ \Dmi\cdot\Sm'\cdot\Emi
\end{equation}
where each estimated value in $\Mm'$ is scaled consistently
with respect to the values in its associated row and column.
For example, suppose that $\Mm(i,j)$ is unknown. It can be 
verified that if row $i$ of $\Mm$ is scaled by a factor $\alpha$,
and column $j$ is scaled by a factor $\beta$, then the value
estimated for element $(i,j)$ will be $\alpha\beta\cdot\Mm'(i,j)$.
In other words, estimated values are scaled
consistently with respect to the implicit units (scale factors)
associated with each row and column, i.e., as associated with
different users and products. 

With regard to computational complexity, the run-time complexity
to determine the RZ scaling is $O(mn)$, which is optimal for 
dense $\Mm$, and the algorithm can be further refined to 
exploit sparsity so that the scaling of $\Mm$ with only $p$ nonzero 
elements can be evaluated in $O(p)$ time\footnote{Optimal
complexity in the sparse limit can justify a simpler model in 
order to permit practical solutions to extremely large 
problems\,\cite{juslam} or in high data-rate 
applications\,\cite{bou}.}. As for space, once the
scaling is determined there is actually no need to explicitly produce 
$\Mm'$ because all that needs to be stored are the diagonal elements 
of $\Dm$ and $\Em$ because the value of each unfilled element 
$(i,j)$ is completely determined as $1/(\Dm_{ii}\cdot\Em_{jj})$. 

The natural question is: {\em Does scale consistency produce 
estimated values that accurately predict user preferences?} The
answer of course can only be assessed by empirical evaluation, which
will be the focus of future work. 
However, a related question can also be asked: {\em Can a
general method be expected to yield accurate predictions across
a range of RS domains if it does \underline{\smash{not}} maintain scale
consistency?} In other words, scale-consistency may not be
sufficient to produce acceptable results, but there is good
reason to believe that it will need to be preserved -- at
least approximately -- unless obviated by the availability of
stronger assumptions about how users produce their 
scores/ratings.

More generally, any RS approach must assume or develop 
a parameterized model for how users determine their 
evaluation scores. The effectiveness with which that model 
actually captures the implicit evaluation formula of a given user 
can be assessed by comparing its predictions against ground 
truth scores from that user. Assuming there will be some users 
who are highly eccentric in the way they determine scores, i.e., 
very different from the implicit assumptions of the model, it makes sense to
identify and remove them so that their inputs don't degrade predictions 
for other users for whom the model appears effective\footnote{The
predictions for deviant/outlier users can be retained as special
cases from the initial calculation while those for the remaining 
users will be evaluated from the more refined scaling obtained
from the regression process.}.

\section{Summary}

In this paper we have described a Recommendation System (RS)
approach that provides consistency with respect to arbitrary
units that are implicitly used by different evaluators when
determining their product scores/ratings. We have provided 
a rationale for why this approach may be effective, and we have
discussed how the accuracy of it (and other RS approaches) potentially
may be improved by a simple regression scheme.
More generally, we believe this approach preserves properties
that are important in a variety of matrix-completion applications.

\end{document}